\documentclass[aps,prl,twocolumn,superscriptaddress,showpacs,draft]{revtex4}
\usepackage{graphicx}
\input{epsf}

\begin{document}

\title{Wigner crystal in snaked nanochannels}

\author{O.V.Zhirov}
\affiliation{\mbox{Budker Institute of Nuclear Physics, 630090 Novosibirsk, Russia}}
\author{D.L.Shepelyansky}
%\homepage[]{http://www.quantware.ups-tlse.fr}
\affiliation{\mbox{Laboratoire de Physique Th\'eorique du CNRS (IRSAMC), 
Universit\'e de Toulouse, UPS, F-31062 Toulouse, France}}
%\affiliation{\mbox{LPT (IRSAMC), CNRS, F-31062 Toulouse, France}}

%\date{\today}
\date{February 7, 2011}

%\pacs{PACS numbers: 05.45.-a, 05.45.Ac, 05.45.Jn}
%\PACS{
%{05.45.-a}{Nonlinear dynamics and chaos}
%\and
%{05.45.Ac}{Low-dimensional chaos}
%{05.45.Jn}{High-dimensional chaos}
%}

%71.45.Lr Charge-density-wave systems
%82.47.Uv Electrochemical capacitors; supercapacitors 
%61.20.Qg Structure of associated liquids: electrolytes, molten salts, etc. 

\pacs{05.45.Ac, 71.45.Lr, 82.47.Uv}
\begin{abstract}
We study properties of Wigner crystal in snaked nanochannels and
show that they are characterized by conducting sliding phase
at low charge densities and insulating pinned phase 
emerging above a certain critical
charge density. The transition between these phases has a devil's staircase 
structure typical for the Aubry transition in  dynamical maps
and the Frenkel-Kontorova model. We discuss implications of this
phenomenon for charge density waves in 
quasi-one-dimensional organic conductors
and for supercapacitors in nanopore materials.
\end{abstract}

\maketitle
\begin{figure}
\centerline{\epsfxsize=8.0cm\epsffile{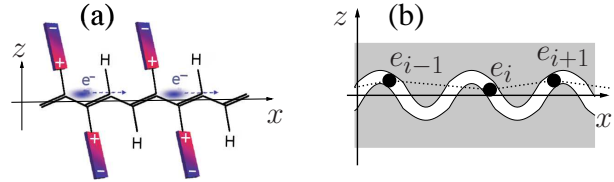}}
\vglue -0.2cm
\caption{(Color online) (a)A schematic image of the Little suggestion
for electron transport in organic molecules (after \cite{little,jerome}).
(b) A schematic image of electron Wigner crystal 
with charges $e_i$ (points) 
sliding in a snaked sinusoidal nanochannel,
dashed lines show force directions between nearby electrons. 
} 
\label{fig1}
\end{figure}

The Wigner crystal \cite{wigner} appears in 
a great variety of physical systems including
electrons in two-dimensional semiconductor samples and 
one-dimensional (1D) nanowires 
(see review \cite{matveev} and Refs. therein),
electrons on a surface of liquid helium \cite{konobook},
cold ions in radio-frequency traps \cite{dubinrmp}
and dusty plasma in laboratory or in space \cite{fortov}.
Effects of Coulomb interactions are clearly seen
experimentally in nanowires \cite{auslaender,ritchie}
and carbon nanotubes \cite{bockrath}.
Also interaction effects for electrons in microchannels
on a surface of liquid helium have been  recently
observed experimentally
\cite{kono2011}. In view of this remarkable  progress
it is interesting to investigate  sliding and
conducting properties of the Wigner crystal 
in wiggled or snaked nanochannels. 
The interest to such studies goes back to the Little suggestion
\cite{little} on electron conduction in long spine
conjugated polymers where he proposed an approach for synthesizing of
organic  superconductors. A modern overview discussion of this 
important suggestion is given in \cite{jerome}. A schematic image
of electron transport in such organic molecules is shown in
Fig.~\ref{fig1}a. According to this picture long molecules form 
wiggling channels which in principle can support 
electron transport along them. However, the Coulomb interactions
between electrons are rather strong 
at such small scales and thus it is not obvious
under what conditions a sliding of Wigner crystal along such channels
is possible. This problem is related 
to the conduction properties of charge density waves
(CDW) (see e.g. reviews \cite{jerome,thorne}). 
To study this phenomenon we choose a simple model of
1D snaked channel shown in Fig.~\ref{fig1}b. There is
no potential gradient along the channel but the channel walls 
are assumed to be very high so  that
electrons can move only along the channel. 

In addition, the properties of 
Wigner crystal  in snaked nanochannels are also useful for
understanding of mechanisms of charge storage in 
electrochemical capacitors, or supercapacitors, which start to
have important industrial applications \cite{gogotsi1,gogotsi2}.
In these systems, charged ions are stored in nanopores 
at the surface of the carbon-activated material
which has enormously large capacitance $C$
going beyond the meanfield values given by 
the Helmholtz theory \cite{shklovskii1,shklovskii2}.
At nanoscale the wiggling of pores is definitely present
that makes our studies very timely.

Due to sinusoidal channel wiggling the Wigner crystal moves 
in a certain effective periodic potential. The case of sliding
of 1D Wigner crystal in a periodic energy potential
was analyzed in \cite{fki} having in mind an example of ion chains in
optical lattices. It was shown there that this problem
can be locally reduced to the Frenkel-Kontorova model
for a particle spring chain in a periodic potential \cite{obraun}
with particle  positions described by the Chirikov standard map 
\cite{chirikov}. For a small amplitude of periodic potential
the Wigner crystal with an incommensurate ion density
can slide in an optical lattice
but above a certain critical amplitude of potential the crystal 
is pinned by the lattice due to the Aubry analyticity breaking
transition \cite{aubry}. In the pinned phase the phonon spectrum
has a gap for long wave excitations so that this regime corresponds to 
an insulating phase. This situation corresponds to a dynamical spin glass
phase with exponentially many stable classical configurations
being exponentially close to a ground state at a fixed 
electron density \cite{fki}. 
The Frenkel-Kontorova model is characterized by similar 
classical and quantum properties \cite{fk1,fk2}.
At sufficiently large values of effective Planck constant 
a quantum instanton tunneling between these quasidegenerate configurations
leads to a zero-temperature quantum phase transition
due to which the quantum Wigner crystal becomes conducting \cite{fki}.
In the following we show that the main elements of this physical
picture remain valid for the Wigner crystal in snaked nanochannels
which are however characterized by enormously sharp
changes of conducting properties.
\begin{figure}
\centerline{\epsfxsize=8.2cm\epsffile{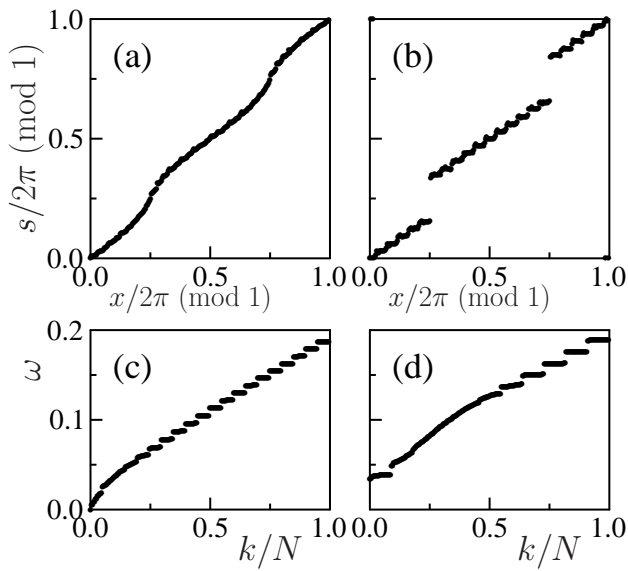}}
\vglue -0.2cm
\caption{(Color online) Hull function $s=h(x)$ (a,b)
and phonon spectrum $\omega(k/N)$ (c,d) 
for incommensurate electron
densities $\nu=N/L=239/233$ (a,c) and $\nu=N/L=244/233$ (b,d).
Here $a=1.2$. 
} 
\label{fig2}
\end{figure}

We start our analysis from  the  case of classical 
electrons with Coulomb interactions moving in a snaked nanochannel
shown in Fig.~\ref{fig1}b. In this case the total system energy $E$ 
is given by a sum  over all Coulomb interactions. Due to 
strong nonlinearity of the system the minimal energy
configurations should be find numerically using the methods
described in \cite{aubry,fki,fk1,fk2}.  We take
a finite number of electrons
$N$ for $L$ periods of a channel of finite length.
In numerical simulations we put the channel on a cylindrical surface
in 3D with electron coordinates being
$x_i = L \sin(s_i/L)$, $y_i = L \cos(s_i/L)$, 
$z=a \sin(s_i)$ where $s_i$ is coordinate along channel
for electron $i$. Thus the channel, filled by $N$ electrons, 
wiggles in $z-$direction
making $L$ periodic oscillations along cylinder of radius $L$
with periodic boundary conditions.
The Coulomb energy of the system is 
\begin{equation}
\label{eq1}
E = \sum_{j>i} 1/R(s_i,s_j)
\end{equation}
where $R(s_i,s_j)$ is the distance between two electrons. We find
from geometry $R^2(s_i,s_j)=4L^2 \sin^2[(s_i-s_j)/2L]+
a^2(\sin s_i - \sin s_j)^2$. Here we choose dimensionless units 
for charge $e$
and length, so that
the channel period  length  is $\ell=2\pi$
and dimensionless amplitude of channel oscillations is $a$.  
In the limit of large $L$ we have a channel wiggling in $(x,z)$ plane.
At $a=0$ we have electrons on a circle.
\begin{figure}
\centerline{\epsfxsize=8.2cm\epsffile{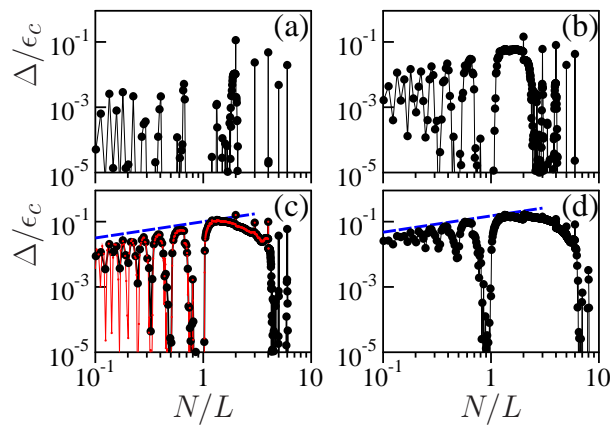}}
\vglue -0.2cm
\caption{(Color online) Dependence of the dimensionless phonon gap 
$\Delta/\epsilon_c$ on  the electron density $\nu=N/L$
for $a=0.7 (a)$, $1 (b)$, $1.2 (c)$, $1.5 (d)$.
Here $L=89$ (black), $233$ (gray/red). The straight line
shows empirical dependence 
$\Delta/\epsilon_c \propto (N/L)^{1/2}$ for 
$(c,d)$, where $\epsilon_c=2\pi e^2\nu/\ell=\nu$ is the Coulomb energy.
} 
\label{fig3} 
\end{figure}

The equilibrium static configurations are defined by the condition
$\partial E/ \partial s_i=0$ with a minimal ground state energy
configuration determined numerically by the standard methods
\cite{aubry,fki,fk1}. Using these methods we also 
find the phonon spectrum $\omega(k)$ of 
small oscillations at the ground state.  
It is easy to see that the total energy $E$
is invariant for a homogeneous shift of all electrons 
by $\delta s$ when the distance between nearby electrons
is $s_{i+1}-s_i = 2\pi m$ that corresponds to electron
density $\nu=N/L$ with
resonant rational values $\nu_m=1/m$.  
Hence, at such a density the Wigner crystal
can freely slide along the channel. For irrational
density values the properties of sliding are 
much more tricky. An example is shown in Fig.~\ref{fig2}
for two very close incommensurate densities $\nu$. For
$\nu=239/233$ we have a smooth hull function
$s_i=h(x_i) (mod 2\pi)$ with the gapless 
phonon spectrum $\omega \propto \omega_0 k/N$ at small wave numbers $k/N$.
Here $x_i (mod 2\pi)$ are ground state electron positions at $a=0$.
The dimensional unit of frequency 
$\omega_0=(e^2/(m (\ell/2\pi)^3))^{1/2}$
is expressed via the particle charge $e$, 
mass $m$ and channel period $\ell$,
we omit it in the further dimensionless computations.
This regime corresponds to the continues 
invariant Kolmogorov-Arnold-Moser (KAM) curves
as it is discussed for the Frenkel-Kontorova model
\cite{fki,obraun,chirikov,aubry,fk1}. Here,
the Wigner crystal can slide freely along the nanochannel.
In contrast, for very close density $\nu=244/233$
the hull function starts to have devil's staircase form,
well known for the cantori regime in the Frenkel-Kontorova model.
Here, the gap $\Delta$ appears in the phonon spectrum 
so that the crystal is pinned in the channel.
This regime corresponds to the insulating phase.

The dependence of phonon gap $\Delta$ on electron density $\nu$
is shown in Fig.~\ref{fig3} for various
values of channel deformation amplitude $a$. At small deformations
the gap is zero for a large fraction of densities $\nu$
(Fig.~\ref{fig3}a) and 
the crystal can slide freely along the channel.
However, at larger deformations
the gap disappears only in a vicinity
of rational densities $\nu_m$ (Fig.~\ref{fig3}b,c)
and at strong deformation regime
only narrow zero gap intervals 
remain around these density values  (Fig.~\ref{fig3}d).
We note that our numerical data are obtained at
rather large number of electrons $N$ and
channel periods $L$ so that
the dependence $\Delta(\nu)$ found numerically
corresponds to the limit of infinite channel length.
Indeed, the function $\Delta(\nu)$
remains practically unchanged with an increase of $L$
(Fig.~\ref{fig3}c). The global dependence of $\Delta$ on $\nu$
corresponds to frequency of small charge oscillations
$\Delta \propto \nu^{3/2} \propto 1/\ell^{3/2}$, being
in agreement with data of Fig.~\ref{fig3}c,d.

A remarkable feature of the dependence
$\Delta(\nu)$ is its very sharp variation with
density $\nu$ and deformation $a$.
The dependence is enormously sharp
in a vicinity $\nu=1$: for $\nu < 1$ there is crystal 
sliding in the channel while only slightly above
$\nu=1$, e.g. for $N/L=1+11/233$, we obtain the insulating phase.
This reminds a sharp change of  conduction properties of 
organic materials with pressure \cite{jerome}
which effectively modifies  $\nu$ and $a$ values.
\begin{figure}
\centerline{\epsfxsize=8.0cm\epsffile{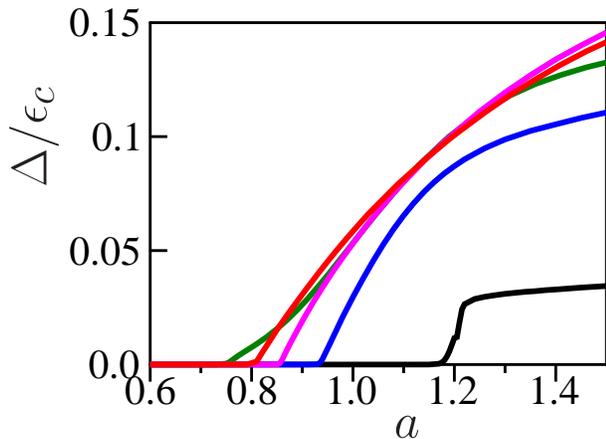}}
\vglue -0.4cm
\caption{(Color online) Dependence of 
rescaled phonon gap $\Delta/\epsilon_c$
on  channel deformation amplitude $a$
at various values of electron density
$\nu$ with the number of electrons
$N=241$ (black), 269(blue), 337 (magenta), 
377 (red),  307 (green) (curves from right to left
at $\Delta/\epsilon_c=0.01$  
respectively)
at $L=233$.
} 
\label{fig4}
\end{figure}

The dependence of phonon gap $\Delta$ on 
channel deformation $a$ is shown in Fig.~\ref{fig4}
for few  density values $\nu$. The gap changes smoothly
with $a$ for $a > a_c$ where $a_c$ is a certain
critical value of deformation which depends of density $\nu$. 
For $a<a_c$ we find very sharp drop of $\Delta$
which becomes exponentially small, e.g. 
$\Delta$ drops by 5 orders of magnitude
when $a$ is decreased by a couple percents in a vicinity of $a_c$.
Since simulations are done at a finite $N$ this means
that in the thermodynamic limit
$\Delta=0$ for $a<a_c$. We interpret
these data in a way similar to the case of the Frenkel-Kontorova
model \cite{aubry,obraun,fk1}: 
for $a<a_c(\nu)$ we have an analytic invariant  KAM
curve with a rotation number corresponding to a given density,
while for $a>a_c$ this curve is replaced by a cantori 
with a finite phonon gap and pinning of the crystal.

\begin{figure}
\centerline{\epsfxsize=8.0cm\epsffile{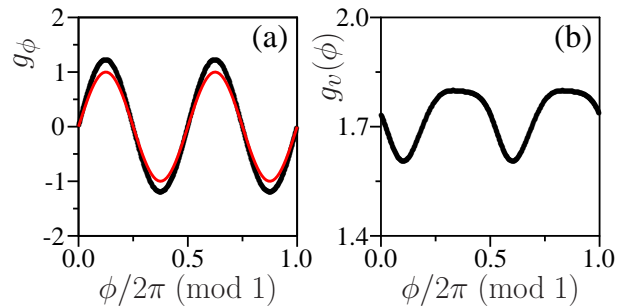}}
\vglue -0.3cm
\caption{(Color online) Map kick functions $g_\phi(\phi)$ (a)
and $g_v(v)$ (b) obtained from the groundstate electron positions
$s_i$ in nanochannel (points), full red/gray curve in (a)
shows the theoretical dependence from the map (\ref{eq2}), see also text.
Here N=377, L=233, $a=0.5$.
} 
\label{fig5}
\end{figure}

To understand in a better way the numerical results presented
above we derive an approximate dynamical map
which determines recursively the electron positions
along the channel. The recursion is given by
equilibrium conditions $\partial E/\partial s_i=0$.
Assuming that $a \ll 1$ we can expand $R$ in $a$
that, after keeping only nearest electron interactions,
gives  recursive relations between
$s_{i-1},s_i,s_{i+1}$. They can be presented in a form of 
dynamical map 
\begin{eqnarray}
\label{eq2}
 {\bar v} &  = & v + 2 a^2 (1-\cos {\bar v}) \sin 2 \phi \; ,\nonumber\\
{\bar \phi} & = & \phi + {\bar v} + a^2 \sin {\bar v} \cos 2 \phi \; ,
\end{eqnarray}
where $v=s_{i}-s_{i-1}$, $\phi=s_i$ are conjugated action-phase variables,
bar marks their values after iteration.
The map is implicit but symplectic (see e.g. \cite{lichtenberg}). 
To check its validity we use the values
$s_i$ obtained for the groundstate configuration
and extract from them the 
kick function $g_\phi=\sin 2\phi$ 
from the values ${\bar v}-v = 2a^2g_v(v) g_\phi(\phi)$
with $g_v(v)=1-\cos v$.
Such a check shows that the map (\ref{eq2})
indeed gives a good description 
of actual electron positions $s_i$
up to moderate values of $a$ (Fig.~\ref{fig5}).

\begin{figure}
\centerline{\epsfxsize=8.0cm\epsffile{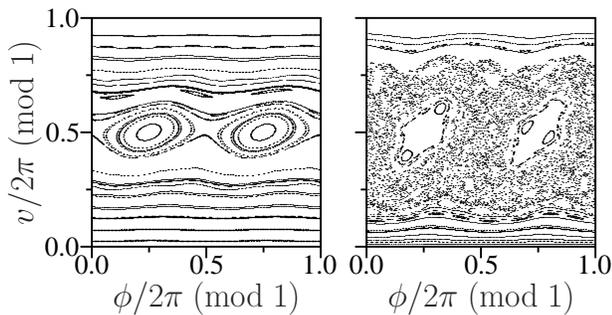}}
\vglue -0.3cm
\caption{(Color online) Poincare section for the dynamical map
(\ref{eq2}) 
at $a=0.25$ (left panel), $0.5$ (right panel).
} 
\label{fig6}
\end{figure}

At small $a$ the phase space of the map
is covered by invariant KAM curves as it is shown in
Fig.~\ref{fig6} (left). At larger $a$ a single chaotic component covers
a significant part of the phase space (Fig.~\ref{fig6} right).
Locally the dynamics is approximately
described by the Chirikov standard map
with the chaos parameter $K \approx 4 a^2 (1-\cos v)$.
According to \cite{chirikov,lichtenberg}
the KAM curves are destroyed at $K>1$ that
is in a good agreement with our numerical data 
of Fig.~\ref{fig3} where the KAM curve
with the golden rotation number
$\nu=0.618...$ goes to the cantori regime 
approximately at $a \approx 0.4$. We note that
at small charge density $\nu$ the parameter
$K$ is small $K \approx 2 a^2 \nu^2 \ll 1$
that corresponds to the KAM regime and a conducting phase
of Wigner crystal in agreement with the data of Fig.~\ref{fig3}.

Of course, the map description is valid only up to moderate
$a$ values. At $a > 1$ the expansion in $a$ 
is no more valid that explains the asymmetry in the
dependence for $\Delta(\nu)$ at $\nu<1$ and $\nu>1$
which is absent in the approximate map (\ref{eq2})
but is clearly seen in Fig.~\ref{fig3}.
Further studies are required to 
obtain a map description at large values of 
deformation $a$.

Our studies determined conditions of sliding and pinning
of the Wigner crystal in snaked nanochannels.
Here, we performed only classical analysis. According to results 
of \cite{fki} 
the quantum effects are weak if the dimensionless
effective Planck constant 
$\hbar_{eff} = (2\pi \hbar^2/m e^2 \ell)^{1/2}$ is small.
This is definitely the case for supercapacitors
with $\ell/2\pi \sim 1 nm$, large ion mass $m \sim 4 \cdot 10^4 m_e$
compared to electron mass $m_e$,  that gives
$\hbar_{eff} \sim 10^{-3}$. The charge storage process in this case
starts with small charge density values $\nu$ where the ions
slide easily in nanochannels since the gap $\Delta$
is practically absent there (see Fig.~\ref{fig3}).
However, with the increase of $\nu$
ions form the Wigner crystal 
which is pinned inside  the nanopores at large 
$\nu$ values.  We think that this is the physical
mechanism behind the charge process of  electrochemical capacitors
studied in \cite{gogotsi1,gogotsi2}. We note that
the energy of pinned Wigner crystal
can be estimated as $W_W \sim  S d e^2/\epsilon (\ell/2\pi)^4 $,
where $S$ is the surface area, $d$ is the deepness
of nanopore layer on
the surface and $\epsilon$ is the dielectric constant. 
For typical parameters $\epsilon = 5$, $\ell/2\pi = 1nm$,
$d = 1 \mu m$ we obtain 
$W_W/S \approx  5 \cdot 10^{-3} J/cm^2$.
It is natural to assume that a part of this energy
can be used during recharging process 
that makes it comparable with the surface
energy density reached in supercapacitors
with $W/S \sim 10^{-3} J/cm^2$ at maximal
capacitance per area $C \approx 400 \mu F/cm^2$
and voltage $U \sim 2 V$ \cite{gogotsi1,gogotsi2}. 
We note that our estimate gives an increase of
$W_W$ with a decrease of nanopore size $\ell$
that qualitatively corresponds to the  behavior observed
experimentally (see e.g. Fig.3a in \cite{gogotsi2}).
At the above parameters the typical pinning frequency
is $\omega_0 \Delta/2\pi \sim 50 GHz$ so that the Wigner crystal should be
sensitive to microwave radiation in this frequency range. 

In contrast, for CDW in organic conductors \cite{jerome}
we have $m \sim m_e$, $\ell/2\pi \sim 3 \AA$ that gives
$\hbar_{eff} \sim 0.5$ so that quantum effects can play
an important role. Further studies are required to analysis
quantum properties of crystal sliding but we expect
that they will have similarities with the quantum Wigner
crystal in a periodic potential \cite{fki}
and the quantum Frenkel-Kontorova model \cite{fk2}.
The classical pinned phase should correspond to the insulator phase,
while we expect that the classical sliding phase may
correspond to the superconducting regime in the quantum case.
Indeed, the sliding phase has a linear dispersion 
law $\omega(k)$ that can be similar to the situation in superfluid phase.
The sharp transitions from conducting to insulating phase 
with charge density variation are well pronounced in 
the classical regime and are expected to be present also
in the quantum case. This can be at the origin of 
high sensitivity of organic conductors to pressure.
Further studies should shed more light on the
quantum properties of Wigner crystal in snaked nanochannels
and organic molecular chains. It would be very interesting to
study such effects experimentally creating
artificial snaked channels with electrons
on a surface of liquid helium \cite{kono2011}.

%
%units: 1Farad= 
%1eV = 1.60217653e-19 J; 1F=1E-9 * c^2 \approx 9e11 cm
%c=light velocity in CGS (Gauss units)
%electron charge e=-4.803E-10(CSG), mass_electron=9.1E-28g
%m_i=40000m_e, ell/2pi=1e-7cm
%1J=1e7 erg

This research is supported in part by ANR PNANO project NANOTERRA.

\vglue -0.3cm

\end{document}